\documentclass[%
 reprint,
runinaddress,
 amsmath,amssymb,
aps,
prb,
showkeys %Show keywords
]{revtex4-1}

\usepackage{graphicx}% Include figure files
\usepackage{hyperref}% Allow interactive references
\usepackage{dcolumn}% Align table columns on decimal point
\usepackage{bm}% bold math
\usepackage[utf8]{inputenc} %Package that allows the Danish letters æ, ø, and å. Neccessary since two of the authors have ø in their last name.
\usepackage{float}
\usepackage{xcolor}

\pdfoutput=1

\newcommand{\dif}{\,\mathrm{d}}
\newcommand{\curl}[1]{\vec\nabla \times #1} % for curl

\renewcommand{\Re}{\text{Re}}			
\renewcommand{\Im}{\text{Im}}

\begin{document}

\bibliographystyle{unsrtnat} %references in numerical order

\title{Quantum spill-out in nanometer-thin gold slabs: Effect on plasmon mode index and plasmonic absorption}% 

\author{Enok J. H. Skjølstrup}
\email{ejs@mp.aau.dk}
\author{Thomas Søndergaard}
\author{Thomas G. Pedersen}
\affiliation{%
 Department of Materials and Production, Aalborg University, Skjernvej 4A, DK-9220 Aalborg East, Denmark. 
}%


\begin{abstract}
A quantum mechanical approach and local response theory are applied to study plasmons propagating in nanometer-thin gold slabs sandwiched between different dielectrics. The metal slab supports two different kinds of modes, classified as long-range and short-range plasmons. Quantum spill-out is found to significantly increase the imaginary part of their mode indices, and, surprisingly, even for slabs wide enough to approach bulk the increase is 20\%. This is explained in terms of enhanced plasmonic absorption, which mainly takes place at narrow peaks located near the slab surface.
\end{abstract}

 %Use showkeys class option if keyword
\maketitle
\section{Introduction} \label{sec:introduction}
Recently, it was found that the effect of quantum spill-out in nanometer-thin gaps in gold has a significant impact on the propagation of surface plasmon polaritons (SPPs) in such structures. In the limit of vanishing gap, the SPP mode index was found to converge to the refractive index of bulk gold, \cite{spill-out} while classical models neglecting spill-out find a diverging mode index \cite{narrow_gap1,gap2}. In addition, it was discovered in Ref. \citenum{spill-out} that spill-out significantly increases the plasmonic absorption in these gaps. Furthermore, the predicted reflectance from an ultrasharp groove array is
in much better agreement with measurements \cite{black_gold} than the classical model. \cite{optics_multiple,GSP}

In this paper, we study the opposite geometry, i.e., a nanometer-thin gold slab surrounded by different dielectrics. Such a structure supports long-range and short-range SPPs, which are $p$-polarized electromagnetic waves bound to and propagating along the slab  \cite{slab1,slab2,slab5,leaking1,slab9}. For a nanometer-thin slab, the short-range mode is strongly bound, meaning that a large part of the field profile is located in the slab region, while the long-range mode is weakly bound with most of its field profile located in the dielectric regions. The magnetic fields of the modes are symmetric and antisymmetric, respectively, if the metal slab is sandwiched between identical dielectrics, while the symmetry is broken when sandwiched between different dielectrics\cite{slab5}. 
Applications of such SPPs are found in, e.g., plasmonic lenses for biosensors and as mode couplers into dielectric or plasmonic waveguides \cite{slab6,slab8}. 
In addition, plasmonic structures find applications within e.g. solar cells\cite{application1}, and furthermore, they can be applied to squeeze light below the diffraction limit \cite{control3,control4}, and can be utilized in lasers\cite{application2}. 

Refs. \citenum{slab1,slab2,leaking1,slab9,slab5} applied a classical model neglecting quantum spill-out, such that the dielectric function takes one value in the metal region and another value in the dielectric region, thus changing abruptly at the interfaces, while quantum effects have been included in, e.g., Refs.\citenum{quantum2,quantum5,quantum6,nonlocal4}.
Furthermore, gold films with thicknesses down to 1 nm have recently been fabricated \cite{maniyara}.

Here, we examine the effect of quantum spill-out on plasmons propagating in nanometer-thin gold slabs. Local response theory is applied to calculate the mode indices and associated electromagnetic fields. We show in the following that spill-out significantly increases the imaginary part of the mode index, even for slabs wide enough to approach bulk. This is explained in terms of strong plasmonic absorption mainly taking place a few Å from the slab surface, a phenomenon not found in classical models.

\section{Quantum dielectric function} \label{sec:quantum}
In the vicinity of the gold slab, the electron density and the effective potential arising from the free electrons (in the $s,p$ band) are significantly modified due to electron tunnelling through the surface barrier. To capture this effect, we calculate the electron density using Density-Functional Theory (DFT) in the jellium model \cite{jellium1,jellium2} (see Appendix A for further description). 
The optical cross sections of metal nanowires \cite{nanowires1,nanowires2,nanowires3,nanowires4}, metal clusters and spheres \cite{clusters1,clusters2,nanowires1,response1} have previously been calculated by applying such a DFT model in the jellium approximation. Likewise the plasmon resonance of metal dimers and semiconductor nanocrystals have been calculated in Refs. \citenum{pol, response2,response3}, while Ref. \citenum{quantum6} studied the plasmonic properties in ultra-thin metal films. 
In addition, Ref. \citenum{abajo} examined the role of electron spill-out and non-local effects on the plasmon dispersion relation for gap plasmons propagating between two gold surfaces as well as plasmons propagating in gold slabs surrounded by air. It was found in that paper, that spill-out has a significant impact while the influence from non-local effects was minor. In Refs. \citenum{quantum6,abajo}, only the real part of the parallel wave number (analogous to mode index) was considered, with no studies of the dependence of the slab (or gap) width. In the present paper, in contrast, we compute both the real and imaginary parts of the mode index, and furthermore, investigate in detail how they depend on the slab width.
Non-local effects in metal dimers and cylinders were studied in Refs.  \citenum{nonlocal5,abajo2,nonlocal1,nonlocal4}, where it was found that these effects slightly blue-shift the plasmon resonances. In this paper, similarly to Refs. \citenum{spill-out,nanowires4,clusters1,clusters2,response3,Ozturk} we ignore the non-local effects and thereby treat the dielectric function as a local response. 

\begin{figure}[!h]
\begin{center}
\includegraphics[width=8cm]{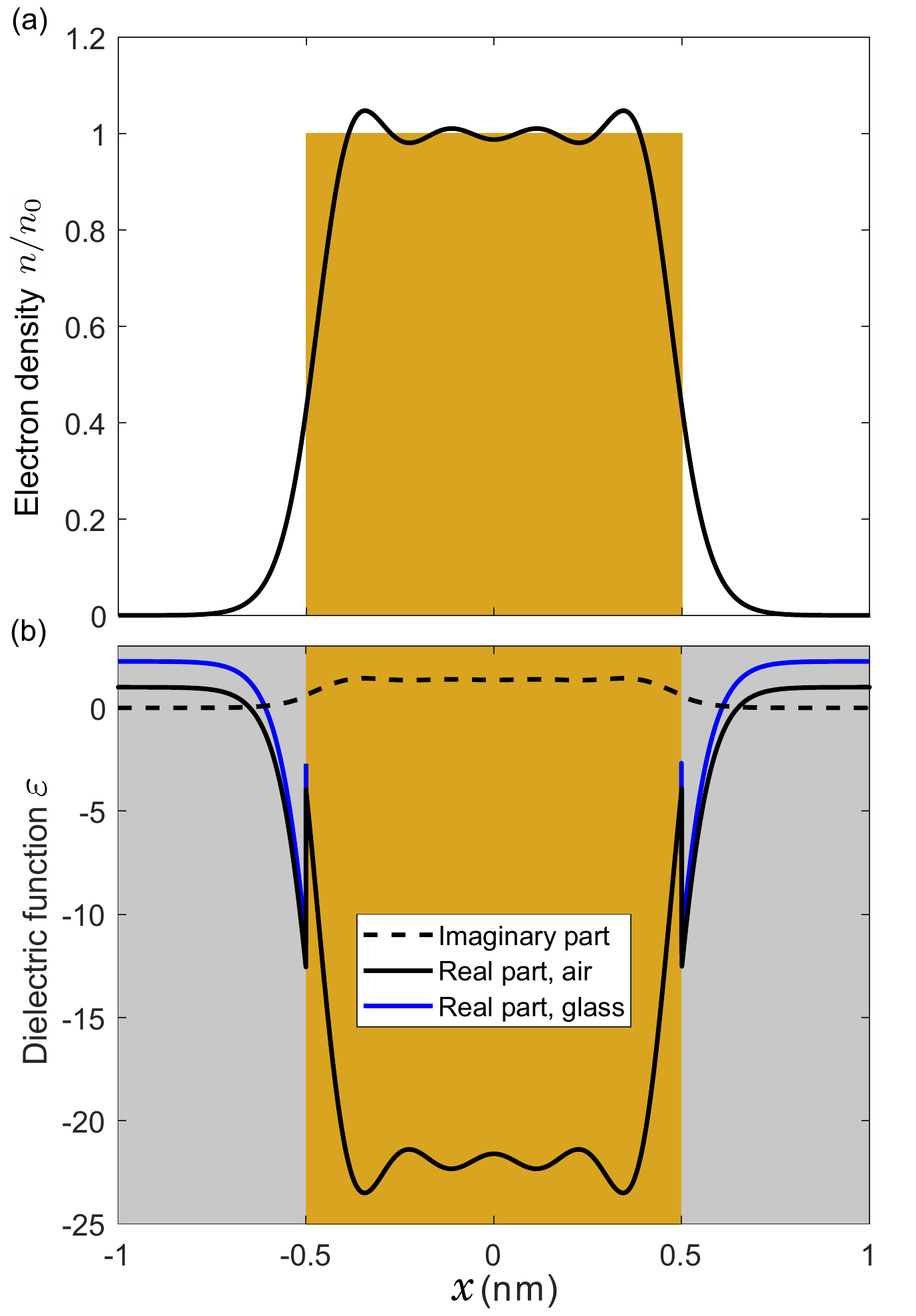}
\end{center}
\caption{(a) Electron density in units of the bulk gold density $n_0$ across a gold slab of width 1 nm. (b) Real and imaginary part of the dielectric function $\varepsilon$ across the same gold slab at a wavelength of 775 nm, where the shaded areas show the position of the surrounding dielectrics. The solid black (blue) curve is for a slab surrounded by air (glass) on both sides. The imaginary part is unaffected by the kind of dielectric. In both figures, the colored areas represent the ion charge.}
\label{fig:density_epsilon}
\end{figure}

The electron density $n$ across a gold slab of width $d=1$ nm is shown in Fig. \ref{fig:density_epsilon}(a) in units of the bulk gold density $n_0$, where the geometry is chosen such that the $x$-axis is perpendicular to the slab, while the plasmons are propagating in the $y$-direction.
The colored area in the figure shows the position of the ion charge in the jellium model, and spill-out is clearly seen to occur as the electron density contains an exponential tail that stretches $\sim$0.3 nm into the dielectric region. In addition, for the charge to be conserved, the density inside the slab is also affected by spill-out. As the slab gets wider, the electron density near the slab boundary contains Friedel oscillations in agreement with Refs. \citenum{jellium1,jellium2,yan} where the electron density at a single interface between gold and air was studied.

The electron density is applied to calculate the local dielectric function $\varepsilon$ across the structure by a method analogous to Ref. \citenum{spill-out}. In the bulk, the electron density $n_0$ implies a bulk plasma frequency of $\omega_{p,\textrm{bulk}}=\sqrt{n_0e^2/(m_e \varepsilon_0)}$, which gives rise to a Drude response $\varepsilon_{p,\textrm{bulk}}(\omega)=1-\omega_{p,\textrm{bulk}}^2/(\omega^2+i\omega\Gamma)$ \cite{nanooptik}. Bound electrons in the lower lying $d$ bands also contribute to the dielectric function\cite{nanooptik}, but in contrast to the free electrons, we assume that they are entirely located in the jellium region, thus not tunnelling through the potential barrier. The response from the bound electrons is calculated from the experimental response of bulk gold, $\varepsilon_{\textrm{gold}}(\omega)$ from Ref. \citenum{christy} as $\varepsilon_{\textrm{bound}}(\omega)=\varepsilon_{\textrm{gold}}(\omega)-\varepsilon_{p,\textrm{bulk}}(\omega)$.
The final dielectric function in the vicinity of a gold slab with a jellium region spanning from $x=-d/2$ to $x=d/2$, is therefore given by
\begin{align}
\varepsilon(\omega,x)&=1-\frac{\omega_p^2(x)}{\omega^2+i\Gamma \omega} + (\varepsilon_{\text{s}}(x)-1)\theta(\vert x \vert-d/2) \nonumber  \\
&+\varepsilon_{\textrm{bound}}(\omega)\theta(d/2-\vert x \vert). \label{eq:eps}
\end{align}
Here, the first term describes the local Drude response of free electrons with position dependent plasma frequency $\omega_p(x)=\sqrt{n(x)e^2/(m_e \varepsilon_0)}$ determined by the electron density $n(x)$ calculated using DFT. Also, $\hbar \Gamma=65.8$ meV has been applied for the damping term\cite{nanooptik}. The dielectric substrate and superstrate, which in general can be different, are described by $\varepsilon_{\text{s}}(x)$, and the Heavyside step function $\theta$ in the first line makes sure that the dielectric function sufficiently far from the slab equals the correct values in the substrate and superstrate. Hence, it has been assumed that the electron density across the slab does not depend on the kind of substrate and superstrate it is surrounded by.
Lastly, the abrupt behaviour assumed for the bound electron term is modelled with the step function $\theta$ in the second line of Eq. \eqref{eq:eps}. 

An example of a dielectric function is seen in Fig. \ref{fig:density_epsilon}(b) for a slab width of 1 nm at a wavelength of 775 nm, where again the colored area shows the position of the ion charge, while the shaded areas depict the dielectric substrate and superstrate. The black curve shows the dielectric function, when there is air on both sides of the slab, while the blue curve corresponds to applying glass ($\varepsilon_s=2.25$) as both substrate and superstrate. 
For a gold slab placed instead on a glass substrate with air as superstrate, the blue curve to the left and the black curve to the right of the slab describe the dielectric function in the glass and air, respectively. The real part of the dielectric function is clearly seen to jump at the slab boundary due to the step function in the second line of Eq. \eqref{eq:eps}. Although it is difficult to see in the figure, the imaginary part of the dielectric function also jumps across the interfaces. Since the substrate and superstrate are assumed lossless, the imaginary part of the dielectric function is unaffected by these materials. 

\section{Mode index of propagating plasmons}
The magnetic field of the SPPs only has a $z$-component and is, for a constant slab width, given by \cite{slab6}
\begin{align}
\vec{H}_m(\vec{r})=\hat{z}H_m(x,y)=\hat{z}\exp(ik_0\beta_my)H_m(x), \label{eq:H}
\end{align}
where the subscripts $m=\{l,s\}$ indicate that the field and associated complex mode index $\beta$ can be either long-range or short-range, respectively, in agreement with Refs. \citenum{slab5}. In Eq. \eqref{eq:H}, $k_0=2\pi/\lambda$ is the free space wavenumber, and $H_m(x)$ is the transverse magnetic field distribution. Both the mode index and the transverse magnetic field depend strongly on $d$, especially for the short-range mode, as will be shown below. 

The mode index is calculated by the same type of transfer matrix method as presented in detail in Ref. \citenum{spill-out} (see Appendix B for classification of modes).
\begin{figure}[!h]
\begin{center}
\includegraphics[width=8cm]{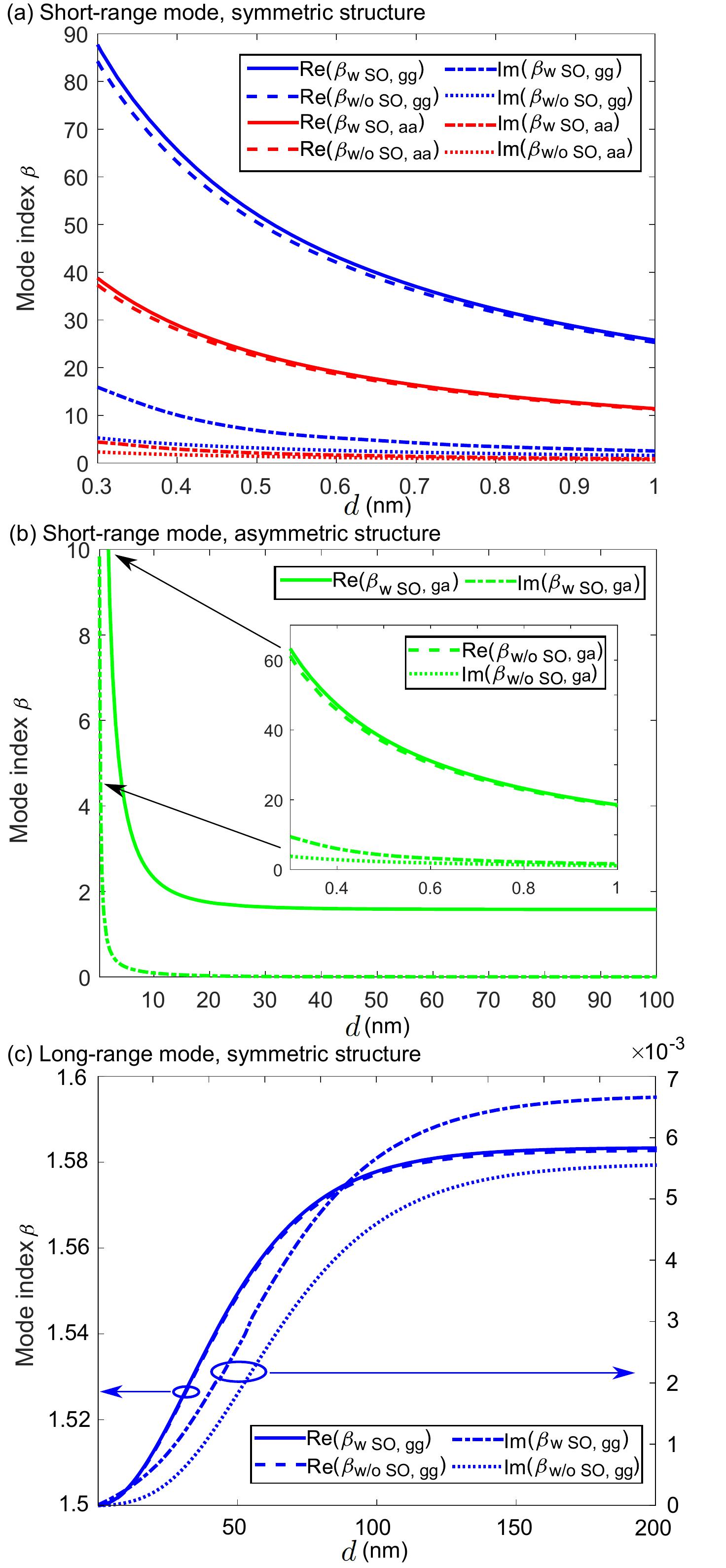}
\end{center}
\caption{Real (solid and dashed lines) and imaginary (dotted and dash-dotted lines) parts of the mode index vs. slab width $d$ at a wavelength of 775 nm. Results are shown for spill-out (SO) included (solid and dash-dotted lines) and neglected (dashed and dotted lines). (a) is for the short-range mode, where the blue and red lines are for slabs surrounded by glass (gg) and air (aa), respectively. (b) is for the short-range mode for a slab surrounded by glass and air (ga), and the inset shows a zoom for $d$ below 1 nm. (c) is for the long-range mode for a slab surrounded by glass (gg), where the real and imaginary parts are shown on the left and right $y$-axes, respectively, as indicated by the arrows. }

\label{fig:modeindex}
\end{figure}
Figure \ref{fig:modeindex} shows the mode indices as a function of $d$ at a wavelength of 775 nm.
For the blue and red curves, the geometric structure is symmetric, while it is asymmetric for the green curves, as indicated in the text above each subfigure. This implies that the magnetic fields associated with the blue and red curves are symmetric and antisymmetric for the long-range and short-range mode, respectively, while the symmetry of the associated fields is broken for the corresponding green curves. For the short-range mode, this will be demonstrated in the next section. 

Figure \ref{fig:modeindex}(a) shows the short-range mode indices for the two symmetric structures, which are in agreement with previous studies \cite{slab1,slab2,slab5} when spill-out is neglected. The real parts are almost unaffected by spill-out, while it plays a significant role for the imaginary parts, as will be elaborated upon below. 
A similar calculation of the mode index with and without spill-out for gap plasmons propagating in narrow gaps in gold showed that the mode index when including spill-out converges to the refractive index of bulk gold in the limit of vanishing gap width\cite{spill-out}, while neglecting spill-out leads to an unphysically diverging mode index \cite{narrow_gap1,gap2}. 
For plasmons bound to the slab, the mode index when neglecting spill-out also diverges unphysically in the limit of vanishing slab thickness\cite{slab1,slab2,slab5}. This is not the case with spill-out included, as plasmonic modes only exist when the real part of the metal dielectric constant is negative in some region along the direction normal to the slab\cite{nanooptik}.
It is found that the slab has to be of sub-atom thickness ($\sim0.3$ Å) in order for the electron density in the model to become so delocalized that the real part of the dielectric constant is everywhere positive.
Hence, with spill-out included the mode index does not diverge in the limit of vanishing slab thickness. Instead, plasmonic modes cease to exist for slabs below a cut-off thickness in the sub-atom range. However, since a gold atom has a diameter of roughly 0.3 nm\cite{black_gold}, we only consider slab widths larger than this value.

For an asymmetric structure with glass as substrate and air as superstrate, the short-range mode index is shown in Fig. \ref{fig:modeindex}(b). Here it is difficult to see the difference in mode index with and without spill-out when the slab width exceeds 1 nm. Therefore the mode index when neglecting spill-out is only included in the inset showing results for $d$ below 1 nm, where $\beta$ has the same behaviour as for the symmetric structure in Fig. \ref{fig:modeindex}(a). Although it is hard to see in the figure, the imaginary part is small but non-zero for all slab widths. 
The mode index has converged when $d=100$ nm, and for such a wide slab the plasmon behaves as if bound to a single interface between glass and gold. \cite{nanooptik} 

The asymmetric structure also supports long-range modes, but only for slab thicknesses above a certain threshold\cite{slab2}. The long-range mode is mainly bound to the air-gold interface, with a mode index that is lower than the refractive index of the glass substrate. This implies that the normal component of the wavevector becomes real (with a very small imaginary part due to loss in the gold) on the glass side of the structure, leading to a wave propagating in the substrate, thus not a truly bound mode \cite{leaking1}. Hence, the wave will leak out into the substrate, where conservation of momentum determines the leakage angle \cite{sondergaard2}. However, if the dielectric constants of the substrate and superstrate are not too different, it is possible to obtain a long-range mode that is truly bound to both interfaces (see e.g. Fig. 3 in Ref. \citenum{slab5}). The phenomenon of leaky modes can be examined using leakage radiation microscopy, see e.g. Refs. \citenum{leaking1,leaking4,leaking5}.
 
Figure \ref{fig:modeindex}(c) shows the mode index of the corresponding long-range mode for a gold slab surrounded by glass. Without spill-out, the mode index for an ultra-thin slab is very close to the refractive index of the substrate, which implies that the mode is weakly bound. The mode is therefore long-range with most of its field profile located in the dielectric regions, which will be illustrated in the next section. As the slab width increases to 200 nm, the mode index without spill-out has converged to $\sqrt{\varepsilon_{\text{gold}}\varepsilon_{\text{glass}}/(\varepsilon_{\text{gold}}+\varepsilon_{\text{glass}})}$, which is the mode index of a plasmon bound to a single interface between gold and glass \cite{nanooptik}. In addition, the corresponding short-range mode index when neglecting spill-out in Fig. \ref{fig:modeindex}(a) converges to the same value for $d=200$ nm (not shown) in agreement with Refs. \citenum{slab2,slab5,slab6}.
With spill-out included, the long-range mode index for small slabs in Fig. \ref{fig:modeindex}(c) is also close to the refractive index of the substrate, and the imaginary part is very low. As for the short-range mode in Fig. \ref{fig:modeindex}(a,b), especially the real part of the mode index is almost the same with and without spill-out, as seen by comparing the solid and dashed lines. But importantly, spill-out significantly increases the imaginary part of the mode index, even for slab widths up to 200 nm, as seen by comparing the dotted and dash-dotted lines in Fig. \ref{fig:modeindex}(c). 

\begin{figure}[!h]
\begin{center}
\includegraphics[width=8cm]{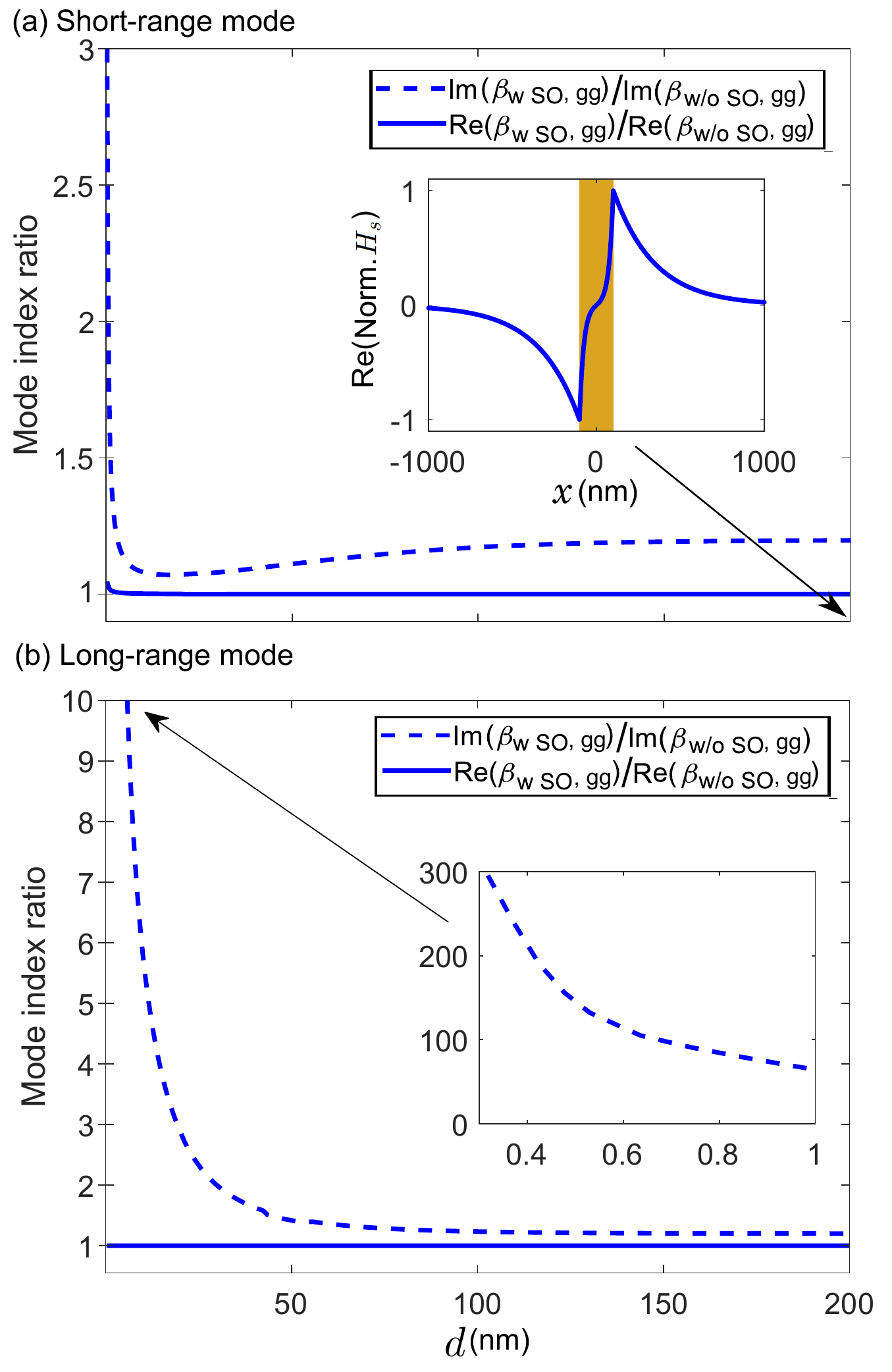}
\end{center}
\caption{Ratio between mode indices with and without spill-out at a wavelength of 775 nm for a slab surrounded by glass (gg). In (a), the mode is short-range, and the inset shows the real part of the transverse magnetic field for $d=200$ nm, as indicated by the arrow, where the colored area represents the ion charge. In (b), the mode is long-range and the inset shows a zoom for $d$ below 1 nm. }
\label{fig:ratio}
\end{figure}

To further illustrate the effect of spill-out in a symmetric structure, the ratio between mode indices with and without spill-out is shown in Fig. \ref{fig:ratio}(a,b) for the short-range and long-range modes, respectively. 
For both types of modes, the real part of the mode index is almost unaffected by spill-out, as the ratios shown by the solid lines in Fig. \ref{fig:ratio} have converged to 1.0004 when $d=50$ nm. For the short-range mode the ratio between the imaginary parts is approximately 3.0 for a slab width of 0.3 nm, while it converges to  $\sim$1.2 for $d=200$ nm. The real part of the normalized magnetic field profile when spill-out is included is shown in the inset of Fig. \ref{fig:ratio}(a) for a slab width of 200 nm, as indicated by the arrow. It behaves as two decoupled plasmons bound to the interfaces between glass and gold, as the field profiles bound to the individual interfaces do not interact for such a wide slab. 

For the long-range mode, the corresponding ratio between the imaginary parts is extremely high for small $d$ as seen in the inset in Fig. \ref{fig:ratio}(b). However, as $d$ increases the ratio decreases monotonically and converges to $\sim$1.2 when $d=200$ nm.  This is an important result showing that quantum spill-out increases the imaginary part of the mode index by 20\%, even for relatively thick slabs that can readily be fabricated \cite{slab6,leaking4} and approach bulk gold. It is highly surprising that spill-out plays such a significant role for wide slabs, as it only modifies the electron density in a region very close to the ion charge. In addition, it is noticed that the ratios between the imaginary parts of the two modes converge to the same value when the slab is wide enough, as in this case the field profiles bound to the individual interfaces are decoupled, similarly to classical models \cite{slab2,slab5,slab6}. Furthermore, the short-range mode indices with and without spill-out in the asymmetric structure in Fig. \ref{fig:modeindex}(b), converge to the same values as for the long-range mode in Fig. \ref{fig:modeindex}(c), as both modes behave as bound to a single interface between gold and glass. Hence, spill-out also increases the imaginary part of the mode index by 20\% in an asymmetric structure. 

Ref. \citenum{quantum6} applied a dielectric function analogous to Eq. \eqref{eq:eps} to study the effect of spill-out on plasmons propagating in a magnesium slab ($r_s=2.66$ Bohr) surrounded by silicon and air. With the present method, the real part of the calculated mode index agrees well with values estimated from Fig. 5 in Ref. \citenum{quantum6}, showing quantitative agreement between that paper and the method presented here. Likewise, the mode index calculated in the present paper agrees well with values estimated from Figs. 6 and S9 in Ref. \citenum{abajo} regarding plasmons propagating in gold slabs surrounded by air.

\section{Field profile and plasmonic absorption} \label{sec:fields}
\begin{figure}[!h]
\begin{center}
\includegraphics[width=8cm]{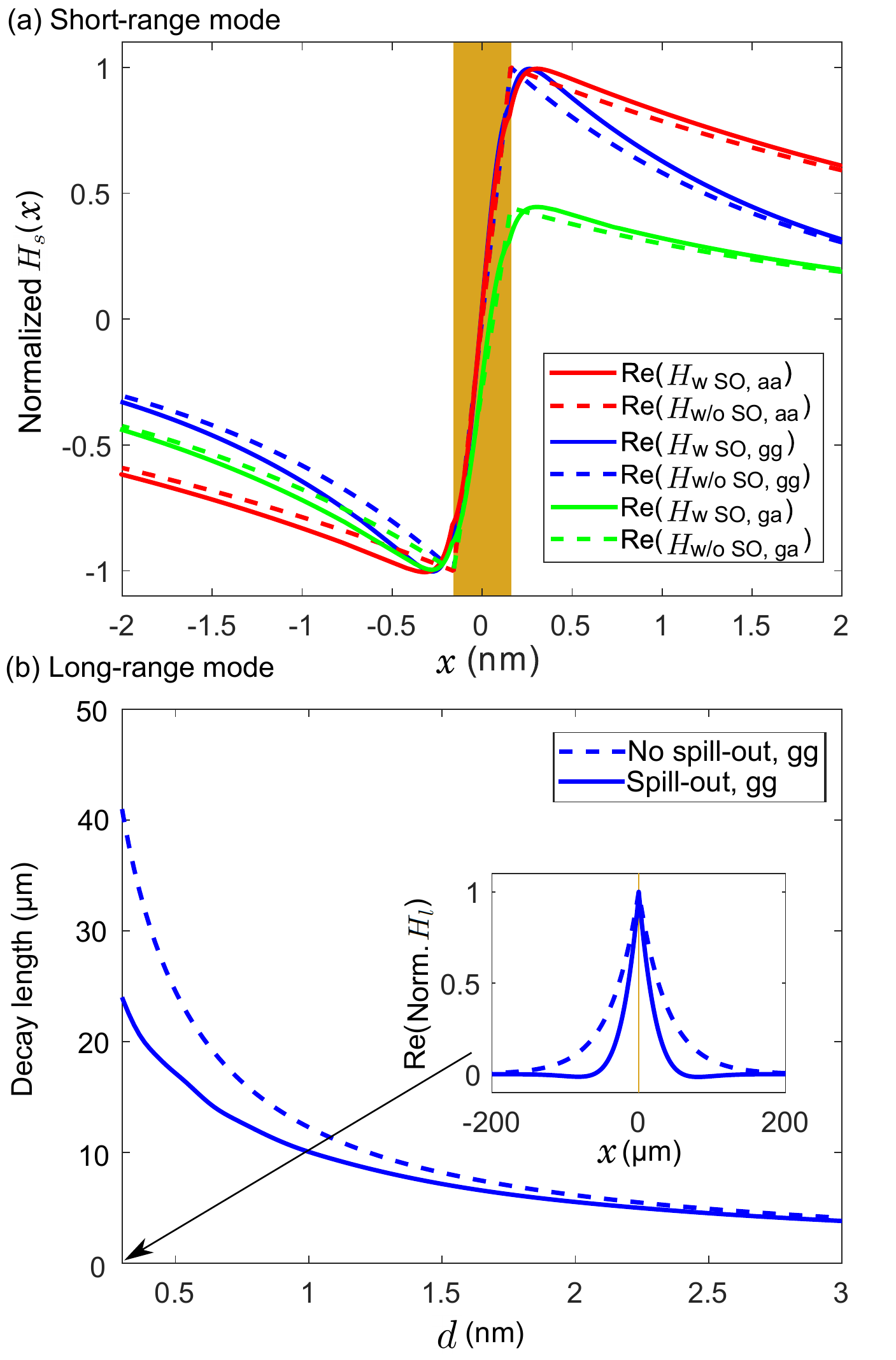}
\end{center}
\caption{(a) Normalized real part of $H_s(x)$ across a slab of width 0.3 nm. (b) shows the decay length of the long-range mode for a gold slab surrounded by glass (gg) with and without spill-out. The inset shows the real part of the magnetic field profiles for a slab of width 0.3 nm as indicated by the arrow. In both (a) and (b), the colored areas represent the ion charge and the wavelength is 775 nm.}
\label{fig:field_penetration}
\end{figure}

Once the mode indices have been computed, the magnetic field from Eq. \eqref{eq:H} is calculated using the same transfer matrix method as described in Refs. \citenum{spill-out,optik}. Applying the same phase convention as in Ref. \citenum{slab1}, the normalized real part of the short-range transverse magnetic field $H_s(x)$ across a gold slab of 0.3 nm is shown in Fig. \ref{fig:field_penetration}(a) at a wavelength of 775 nm.
The associated imaginary parts of the fields are not shown as they are small compared to the real parts, similarly to classical models \cite{slab1}. 

When neglecting spill-out, the slope of the magnetic field, corresponding to the normal component of the electric field, becomes discontinuous across the slab surfaces in agreement with Refs. \citenum{slab1,slab2,slab5}. With spill-out included, the slope is still discontinuous due to the abrupt change in the bound electron term in Eq. \eqref{eq:eps}, although it is difficilt to see in Fig. \ref{fig:field_penetration}(a). But in the vicinity of the slab surface the field profiles behave more smoothly, and their maximum positions are slightly shifted into the dielectric region. We have checked that the appropriate boundary conditions regarding electromagnetic fields across an interface \cite{nanooptik} are satisfied. Further away from the slab, the field profiles with and without spill-out become almost identical. Consequently, the decay lengths into the dielectrics, calculated as $1/\Im{(k_x)}$, where $k_x=k_0\sqrt{\varepsilon_s-\beta^2}$ is the wavenumber in the $x$-direction, are very similay and both are on the order of a few nm. This illustrates that the short-range mode is strongly bound to the slab, as it decays very rapidly into the dielectrics \cite{slab5}, and thereby has a large part of its field profile located in the slab region. Notice, that as the real part of the mode index is much higher than its imaginary part, the decay length mostly depend on the real part of the mode index. 
Fig. \ref{fig:field_penetration}(a) demonstrates that the short-range magnetic field is antisymmetric for the two symmetric structures shown by the blue and red curves, while this is no longer the case for an asymmetric structure, as shown by the green curves. When the slab width increases, the field profile broadens, as shown in the inset of Fig. \ref{fig:ratio}(a) for a 200 nm wide slab. For such a wide slab, the field profiles with and without spill-out are almost identical, and both behave as two decoupled plasmons bound to the interfaces between glass and gold. 

As mentioned above, the long-range mode is weakly bound. Consequently, the electromagnetic fields for a few-nm slab have decay lengths of several micrometers, as shown for a gold slab surrounded by glass in Fig. \ref{fig:field_penetration}(b). The long decay length implies that most of the field profiles are located in the dielectric regions. The field profiles are broader when spill-out is neglected, as also seen in the inset showing the real part of the magnetic fields across a slab of width 0.3 nm, i.e., the same slab as in Fig. \ref{fig:field_penetration}(a). Including spill-out effectively implies a broader slab (see Fig. \ref{fig:density_epsilon}(a)), which means that the fields become slightly more localized with a shorter decay length. For slabs wider than 3 nm, the decay lengths with and without spill-out are very similar, and both converge to the decay length of a plasmon bound to a single interface between gold and glass (not shown).

\begin{figure}[!h]
\begin{center}
\includegraphics[width=8cm]{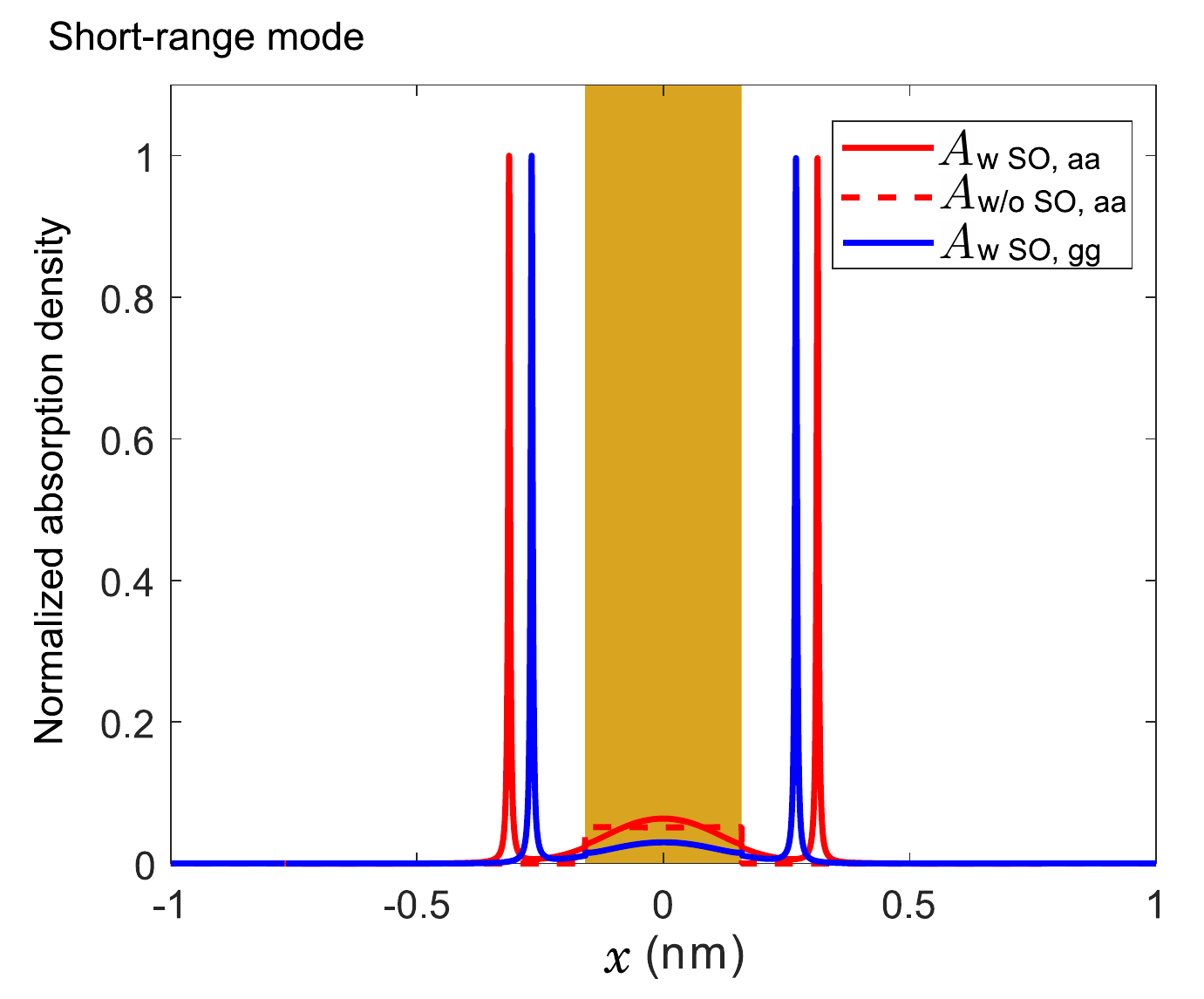}
\end{center}
\caption{Normalized absorption density for the short-range mode across a gold slab of width 0.3 nm at a wavelength of 775 nm. The solid and dashed lines show the absorption density when spill-out is included and neglected, respectively, where the red and blue lines are for slabs surrounded by air (aa) and glass (gg), respectively. }
\label{fig:abs}
\end{figure}

As argued above, spill-out plays almost no role for the decay length for slabs of a few nm. On the other hand, it significantly increases the imaginary part of the mode index as shown in Fig. \ref{fig:ratio}. This leads us to investigate how spill-out affects the electric field and plasmonic absorption across the slab. First, the electric field is calculated from the magnetic field in Eq. \eqref{eq:H} as \cite{nanooptik}
\begin{align}
\vec{E}_{m}(x,y)=\frac{i}{\omega \varepsilon_0\varepsilon(x,y)}\curl \big[\hat{z}H_{m}(x,y)\big]. \label{eq:E}
\end{align}
The electric field is subsequently used to calculate the absorption density defined as
\begin{align}
A_{m}(x,y)=\vert \vec{E}_{m}(x,y)\vert^2 \Im{(\varepsilon(x,y))}. \label{eq:abs}
\end{align}
By considering the time average of the Poynting vector, $\langle \vec{S}\rangle=1/2\Re (\vec{E}\times \vec{H}^*)$\cite{nanooptik}, it can be shown that conservation of energy implies that the plasmonic absorption and the imaginary part of the mode index are related in the following way
\begin{align}
\Im(\beta)=\frac{c\varepsilon_0\int A_m(x,y)\dif x}{2\int \Re (\vec{E}_m(x,y)\times \vec{H}_m^*(x,y))\cdot \hat{y}\dif x}.  \label{eq:relation}
\end{align}
We have checked that this relation is satisfied for both types of modes with and without spill-out.
The normalized absorption density is shown in Fig. \ref{fig:abs} for the short-range mode across the slab of width 0.3 nm at a wavelength of 775 nm.
If spill-out is neglected, absorption can only take place in the gold as the surrounding dielectrics are assumed lossless. In this case, the absorption density is almost unaffected by the kind of surrounding dielectric, why Fig. \ref{fig:abs} only shows it for a slab surrounded by air. But with spill-out included, strong plasmonic absorption occurs, and the absorption density mostly consists of two narrow peaks located in the dielectric regions close to the interfaces. At these positions, similarly to Refs. \citenum{spill-out,Ozturk}, the real part of the dielectric function is zero (at the wavelength 775 nm), while its imaginary part is small but nonzero, which ensures that the peaks in the absorption density are finite. The narrow peaks are found a few Å outside the ion charge, and the same is found for the long-range mode (not shown). The contribution from these peaks leads to enhanced plasmonic absorption, as they are a consequence of electron spill-out, and therefore not found in classical models. 
For a slab surrounded by glass, the peaks occur slightly closer to the slab, as the real part of the dielectric function has its zero shifted slightly compared to the case with a slab surrounded by air (see Fig. \ref{fig:density_epsilon}(b)). The same phenomenon was found in Ref. \citenum{quantum6} for a magnesium slab surrounded by silicon and air. The peaks in absorption density due to spill-out were recently discussed in Ref. \citenum{spill-out}, where they were found to significantly reduce the reflectance from an ultrasharp groove array in much better agreement with measurements\cite{black_gold} compared to classical models \cite{GSP,optics_multiple}.

The increased absorption loss due to spill-out will manifest itself as decreased propagation lengths in fabricated plasmonic structures. Losses in such structures have been studied in, e.g., Refs. \citenum{losses1,losses2,losses3,losses4}, where it was found that the measured propagation length of plasmons propagating in a 70 nm silver film deposited on glass is significantly shorter than the one calculated using classical models\cite{losses3,losses4}. Hence, these works together with Ref. \citenum{spill-out} also support the finding that a classical model is not sufficient to correctly describe losses occurring in plasmonic waveguides.

\section{Conclusion} \label{sec:conclusion}
In conclusion, we have applied a quantum mechanical approach and local response theory to study the propagation of plasmons in nanometer-thin gold slabs surrounded by different dielectrics. 
The effect of spill-out is found to be small on the real part of the mode indices but remarkably increases the corresponding imaginary part, and even for slabs wide enough to approach bulk the increase is 20\%. This is explained in terms of enhanced plasmonic absorption mainly taking place at narrow peaks located a few Å outside the ion charge. It is highly surprising that spill-out plays such a significant role for wide slabs, as it only modifies the electron density in a region very close to the ion charge. For slab widths above a few nanometer, the decay length of the fields into the dielectrics is almost unaffected by spill-out, as it mostly depends on the real part of the mode index.
Furthermore, in contrast to classical models, the short-range mode index does not diverge in the limit of vanishing slab thickness when spill-out is included. Instead, plasmonic modes cease to exist for slab widths below a cut-off thickness in the sub-atom region.

\section*{Acknowledgement}
This work is supported by Villum Kann Rasmussen (VKR) center of excellence Quscope. 

\section*{Appendix A: Calculation of electron density}
In this appendix, we discuss in more detail how the electron density is calculated. 
Within the jellium model it is assumed that the charge of the gold ions is smeared out, such that their charge density is constant within the slab\cite{jellium1,jellium2}. The characteristic spill-out, as seen in Fig. \ref{fig:density_epsilon}(a), stems from the distribution of free electrons in the vicinity of this positive background. The Kohn-Sham equations \cite{jellium1} are solved self-consistently within the local density approximation (LDA), \cite{kohanoff} applying the Perdew-Zunger parametrization \cite{perdew} for the correlation term. The applied Wigner-Seitz radius for gold is $r_s=3.01$ Bohr\cite{jellium2}.

It is found that 2,500 basis functions on the form $\sin(m\pi(x/L+1/2))$ are sufficient to describe the density for slab widths up to 200 nm. The length $L$ is 1 nm larger than the slab width $d$, and the slab is centered at $x=0$. As in Ref. \citenum{spill-out}, the density is said to converge when a variation in Fermi energy between two iterations below $10^{-7}$ Ha is achieved. Furthermore, in the Anderson mixing scheme \cite{anderson}, the mixing parameter $\alpha$ must be below a certain threshold which strongly decreases with $d$. It is found that $\alpha\leq 5\cdot 10^{-4}$ is necessary for slab thicknesses up to 20 nm. The potentials for wider slabs can afterwards be constructed from the potential of the 20 nm slab, as the oscillations in potential near its center are negligible, meaning that the effective potential near the center can be seen as constant. This constant potential is added in the central region of wider slabs. 

\section*{Appendix B: Classification of plasmonic modes}
In this short appendix, we discuss how the plasmonic modes are classified. The mode index is calculated by the same type of transfer matrix method as presented in detail in Ref. \citenum{spill-out}. However, only one type of plasmonic mode was studied in that paper, why the classification of modes was not presented there. A structure matrix $\mathcal{S}$ is constructed, which relates the magnetic fields to the left and right of the structure, and a mode index is found when the matrix element $\mathcal{S}_{11}$ is zero. The mode is classified by the sign of $\mathcal{S}_{21}$, where positive and negative signs correspond to long-range and short-range modes, respectively, and $\mathcal{S}_{21}$ is exactly $\pm 1$ for symmetric structures. Expressions for the matrix elements can be found in Refs. \citenum{spill-out,optik}. In addition, the $x$-axis is divided into tiny segments, each modelled as having a constant dielectric function. Similarly to Ref. \citenum{spill-out}, we find that segments of $2.7\cdot 10^{-4}$ nm are sufficient to avoid discretization errors.

\bibliography{mybib} 

\end{document}